\documentstyle[11pt]{article}
\begin{document}
\title{{\bf Photopion reactions on deltas preexisting in nuclei}}
\author{A.\,Fix\\
{\it Tomsk Polytechnic University, 634004 Tomsk, Russia}\\
I.\,Glavanakov,\ Yu.\,Krechetov\\
{\it Nuclear Physics Institute, 634050 Tomsk, Russia}}
\date{\today}
\maketitle

\large
\begin{abstract}
Reactions $A(\gamma,\pi^+p)$ are considered to proceed through the
formation of $\pi^+p$ pairs on $\Delta$ constituents
in nuclei. We develop the nonrelativistic operator
of $\gamma\Delta^{++}\to\pi^+p$ process in an arbitrary frame.
The calculated cross section of $^{12}C(\gamma,\pi^+p)$ reaction
is compared to the existing experimental data.
\end{abstract}
\bigskip

The role of isobars as nuclear constituents was a topic of a large
body of research over a long time (a comprehensive review can be found
in Refs.\cite {Green76,Aren78}). There is a convincing evidence
that a small admixture of enternally excited nucleons always presents
in nuclei.
Since these virtual excitations, proceeding during the collisions
$NN\to N^*N(N^*N^*)$, are more intensive at the large relative momenta
of nucleons, the isobar configurations are assumed to be
responsible for high momentum components of nuclear wave functions.
Therefore, they can manifest themselves in nuclear reactions
with large momentum transfers.

However, more direct way to demonstrate the existence of isobar
constituents in nuclei as well as to gain a better insight into their
dynamical properties is to study the so-called isobar
knock-out proceses \cite{Gera69}.
In this regard, the $A(\gamma,\pi^+p)$ reactions,
which can be interpreted merely as absorption of incoming photon on
$\Delta^{++}$ preformed in a target nucleus followed by its
real decay into $\pi^+p$ pair, are the most attractive ones.
The main advantage of this channel is that it is free of
any influence of direct photoproduction mechanism, in contrast to the
$A(\gamma,\pi^+n)$ reaction, where the contribution of a preexisting
$\Delta^+$ will be overwhelmed by the $\Delta^+$ creation on a quasifree
proton in $\gamma p\to \Delta^+\to \pi^+n$ process.

To date, only one exclusive measurement of $A(\gamma,\pi^+p)$
reaction has been reported in Ref.\cite{McKenz97}, where the results
of the first comparative studies of $^{12}C(\gamma,\pi^+n)$ and
$^{12}C(\gamma,\pi^+p)$ channels are presented.
While the $^{12}C(\gamma,\pi^+n)$ cross section is dominated
by the quasifree mechanism, the higher order FSI (Final State Interaction)
processes involving also charge exchange are assumed
to be the source of the $^{12}C(\gamma,\pi^+p)$ reaction yield.
The latter assumption is, however, not undoubtedly confirmed by a
calculation performed within the Valencia model (VM) \cite{Oset92},
which markedly underestimates the data.

In this paper we want to explain the observed $^{12}C(\gamma,\pi^+p)$
cross section, at least partially, as a signature of the $\Delta^{++}$
admixture in $^{12}C$ ground state.

The starting point of our discussion is the elementary $\gamma\Delta^{++}
\to\pi^+p$ amplitude, which we need as input for a nuclear calculation.
It is schematically shown in Fig.1, where
the energies and the momenta of participating particles are also labeled.
The matrix element is constructed in analogy with one-particle exchange
model. The relevant diagrams are depicted in Fig.1(a-d).
We restrict
ourselves to nonrelativistic limit keeping only terms up to the order
$(p/M)^2$. The $\pi N\Delta$ vertex is
\begin{equation}
\Gamma_{\pi N\Delta}=-i\displaystyle\frac{f_\Delta }{m_\pi}F_\pi(q_\mu^2)
\,(\vec{\sigma}_{N\Delta}\cdot\vec{q}_{\pi N}\:)\,,
\end{equation}
where $\vec{q}_{\pi N}=(\vec{q}M_N-\vec{p}_NE_\pi)/(M_N+E_\pi)$
is the relative momentum in $\pi N$ system.
The transition spin operator $\vec{\sigma}_{N\Delta}$
is defined by its reduced matrix element
\begin{equation}
\langle\frac{1}{2}\parallel\sigma_{N\Delta}\parallel\frac{3}{2}\rangle
=-2\,.
\end{equation}
For the $\pi N\Delta$ coupling constant we take the value
$f_\Delta^{\:2}/4\pi$\,=\,0.37 obtained from the $\Delta$ decay.
It is generalized to the off-shell region by the conventional monopole
form factor
\begin{equation}
F_\pi(q_\mu^2)=\displaystyle\frac{\Lambda^2-m_\pi^2}{\Lambda^2-q_\mu^2}
\end{equation}
with a cut-off mass $\Lambda$\,=\,1 GeV and $q_\mu$ being the pion
4-momentum.

For the nucleon and isobar currents the following expressions are used
\begin{equation}
\vec{j}_N=\frac{e}{2M_N}\left[e_N(\vec{p}_N+\vec{p}_N^{\:\prime})
+\mu_N\,i(\vec{\sigma}_N\times\vec{k}\,)\right]\,,
\end{equation}
\begin{equation}
\vec{j}_\Delta=\frac{e}{2M_\Delta}\left[e_\Delta(\vec{p}_\Delta+
\vec{p}^{\:\prime}_\Delta)
+\mu_\Delta\,i(\vec{\sigma}_\Delta\times\vec{k}\,)\right]\,.
\end{equation}
Here $\vec{p}_N\,(\vec{p}_\Delta)$ and
$\vec{p}_N^{\:\prime}(\vec{p}^{\:\prime}_\Delta)$ are the incoming
and outgoing nucleon (isobar) momenta, respectively. The charge and the
magnetic moment of proton and $\Delta^{++}$ are
\begin{equation}\label{emu}
e_p=1\,, \quad \mu_p=2.79\,\mbox{n.m.}\,, \quad e_{\Delta^{++}}=2\,,
\quad \mu_{\Delta^{++}}=-2.3\,\mbox{n.m.}\,,
\end{equation}
where the value of $\mu_{\Delta^{++}}$ was obtained in \cite{Chao74}
within a bootstrap model.
The spin operators $\vec{\sigma}_N$ and $\vec{\sigma}_\Delta$
have the following normalization
\begin{equation}
\langle\frac{1}{2}\parallel\sigma_N\parallel\frac{1}{2}\rangle
=\sqrt{6}, \quad
\langle\frac{3}{2}\parallel\sigma_\Delta\parallel\frac{3}{2}\rangle
=2\sqrt{15}\,.
\end{equation}
After taking together all terms depicted in
Fig.1(a-d), it is convenient to split the resulting amplitude
into the vector and the tensor part
\begin{equation}\label{AA}
t_\lambda=i(\vec{A}\cdot\vec{\sigma}_{N\Delta})+
\sqrt{5}\left\{A^{[2]}\otimes\sigma_{N\Delta}^{[2]}
\right\}^{[0]}\,,
\end{equation}
where the spin operator
\begin{equation}
\sigma_{N\Delta}^{[2]}\equiv
\left\{\sigma_N^{[1]}\otimes\sigma_{N\Delta}^{[1]}\right\}^{[2]}
\end{equation}
has the reduced matrix element
\begin{equation}
\langle\frac{1}{2}\parallel\sigma_{N\Delta}^{[2]}
\parallel\frac{3}{2}\rangle=-\sqrt{10}\,.
\end{equation}
The quantities $\vec{A}$ and $A^{[2]}_\mu$ defined by (\ref{AA})
are functions of $\vec{p}_N$ and $\vec{p}_\Delta$ as well as the photon
polarization vector $\varepsilon_\lambda$ ($\lambda$=$\pm$1).
After straightforward manipulation we find
\begin{eqnarray}
\vec{A}&=&-\sqrt{2}\:\frac{ef_\Delta}{m_\pi}
\Bigg\{
\vec{\varepsilon}_\lambda +
F_\pi(t)\frac{2(\vec{q}-\vec{k})(\vec{q}\cdot\vec{\varepsilon}_\lambda)}
{t-m_\pi^2}
\nonumber\\
&+&\vec{q}_{\pi N}\left(
\frac{e_\Delta(\vec{p}_\Delta\cdot\vec{\varepsilon}_\lambda)}
{M_\Delta(E_\Delta+E_\gamma-E_\Delta^{\prime}+\frac{i}{2}\Gamma_\Delta)}
+\frac{e_N(\vec{p}_N\cdot\vec{\varepsilon}_\lambda)}
{M_N(E_\Delta-E_\pi-E_N^{\prime})}\right)\\
&-&
\left[\vec{q}_{\pi N}\times[\vec{k}\times\vec{\varepsilon}_\lambda]\right]
\left(\frac{5\mu_\Delta}{4M_\Delta(E_\Delta+E_\gamma-E_\Delta^{\prime}
+\frac{i}{2}\Gamma_\Delta)}
+\frac{\mu_N}{4M_N(E_\Delta-E_\pi-E_N^{\prime})}\right)
\Bigg\}
\nonumber\\
&&\nonumber\\
A^{[2]}_\mu &=& \sqrt{2}\:\frac{ef_\Delta}{m_\pi}\left\{
[\vec{k}\times\vec{\varepsilon}_\lambda]^{[1]}\otimes
q_{\pi N}^{[1]}\right\}^{[2]}_\mu
\nonumber\\
&&\phantom{xxxxxxx}\times
\left(\frac{\mu_\Delta}{2M_\Delta(E_\Delta+E_\gamma-E_\Delta^{\prime}
+\frac{i}{2}\Gamma_\Delta)}
+\frac{\mu_N}{2M_N(E_\Delta-E_\pi-E_N^{\prime})}\right)\,.
\label{A2}
\end{eqnarray}
Here
$E_N^{\prime}=M_N+(\vec{p}_\Delta-\vec{q}\,)^2/2M_N$ and
$E_\Delta^{\prime}=M_\Delta+(\vec{p}_\Delta+\vec{k}\,)^2/2M_\Delta$
are the energies of intermediate nucleon and delta
and $t=(E_\gamma-E_\pi)^2-(\vec{k}-\vec{q}\,)^2$.
We use the energy dependent resonance width
\begin{equation}
\Gamma_\Delta(W)=
115\left(\frac{q_{\pi N}}{q_\Delta}\right)^3\frac{M_\Delta}{W}\:
\mbox{MeV}\,,
\end{equation}
with $q_\Delta$\,=\,1.64\,$m_\pi$ and $W$ being the $\pi N$ invariant mass.

If we restrict ourselves to the c.m.\ system,
the formal structure of the amplitude becomes
similar to that for $\gamma N\to\pi N$ process written in
CGLN form \cite{CGLN}
\begin{eqnarray}
t_\lambda &=&
iF_1(\vec{\sigma}_{N\Delta}\cdot\vec{\varepsilon}_\lambda)
+F_2(\vec{\sigma}_N\cdot[\hat{k}\times\vec{\varepsilon}_\lambda])
(\vec{\sigma}_{N\Delta}\cdot\hat{q})
\nonumber\\&&\phantom
{iF_1(\vec{\sigma}_{N\Delta}\cdot\vec{\varepsilon}_\lambda)}
+iF_3(\vec{\sigma}_{N\Delta}\cdot\hat{k})(\hat{q}\cdot
\vec{\varepsilon}_\lambda)
+iF_4(\vec{\sigma}_{N\Delta}\cdot\hat{q})(\hat{q}\cdot
\vec{\varepsilon}_\lambda)\,,
\end{eqnarray}
with $\hat{p}\equiv\vec{p}/p$.
The amplitudes $F_i$ read
\begin{eqnarray}
F_1&=&-\sqrt{2}\:\frac{ef_\Delta}{m_\pi}
\left[1+\frac{\mu_\Delta(\vec{k}
\cdot\vec{q})}{M_{\Delta}(E_\Delta+E_\gamma-E_\Delta^{\prime}
+\frac{i}{2}\Gamma_\Delta)}\right]\,,
\nonumber\\
F_2&=&\sqrt{2}\:\frac{ef_\Delta}{m_\pi}kq
\left[\frac{\mu_\Delta}
{2M_{\Delta}(E_\Delta+E_\gamma-E_\Delta^{\prime}+\frac{i}{2}\Gamma_\Delta)}
+\frac{\mu_N}{2M_N(E_\Delta-E_\pi-E_N^{\prime})}\right]\,,
\nonumber\\
F_3&=&\sqrt{2}\:\frac{ef_\Delta}{m_\pi}kq
\left[\frac{2F_\pi(t)}
{t-m_\pi^2}+\frac{\mu_\Delta}
{2M_\Delta(E_\Delta+E_\gamma-E_\Delta^{\prime}
+\frac{i}{2}\Gamma_\Delta)}\right]\,,
\\
F_4&=&-\sqrt{2}\:\frac{ef_\Delta}{m_\pi}q^2
\left[\frac{2F_\pi(t)}
{t-m_\pi^2}-\frac{e_N}
{2M_N(E_\Delta-E_\pi-E_N^{\prime})}\right]\,.
\nonumber
\end{eqnarray}
Finally, the unpolarized differential c.m.\ cross section is
related to the matrix element by
\begin{equation}
\frac{d\sigma}{d\Omega}=\frac{q}{k}\,\frac{M_\Delta M_N}{(4\pi)^2W^2}
\,\frac{1}{2}\sum_{\lambda=\pm 1}\overline{|t_\lambda|}^2\,,
\end{equation}
where
\begin{equation}\label{sig}
\overline{|t_\lambda|}^2=\frac{1}{4}\sum_{m_\Delta m}
|\langle\frac{1}{2}m|t_\lambda|\frac{3}{2}m_\Delta\rangle|^2=
\frac{1}{3}|\vec{A}|^2+\frac{1}{2}\sum_\mu|A^{[2]}_\mu|^2\,.
\end{equation}
Using the explicit expression for $A^{[2]}_\mu$ (\ref{A2}), the last term in
(\ref{sig}) may be written as
\begin{eqnarray}
\sum_\mu|A^{[2]}_\mu|^2&=&
2\left(\frac{ef_\Delta}{m_\pi}\right)^2
\left[\frac{\mu_\Delta}
{2M_{\Delta}(E_\Delta+E_\gamma-E_\Delta^{\prime}+\frac{i}{2}\Gamma_\Delta))}
+\frac{\mu_N}{2M_N(E_\Delta-E_\pi-E_N^{\prime})}\right]^2
\nonumber\\
&\times&\left[k^2q_{\pi N}^2
-\frac{1}{2}k^2(\vec{q}_{\pi N}\cdot\vec{\varepsilon}_\lambda\,)^2
-\frac{1}{2}(\vec{q}_{\pi N}\cdot\vec{k}\,)^2-\frac{1}{3}(\vec{q}_{\pi N}
\cdot[\vec{k}\times\vec{\varepsilon}_\lambda]\,)^2\right]\,.
\end{eqnarray}

The calculated $\gamma\Delta^{++}\to\pi^+p$ cross section
is presented in Fig.2,
where in addition the separate contributions
from the various terms depicted in Fig.1 are also shown.
The distinguishing feature of the differential
cross section (Fig.2a)
is its localization in the backward hemisphere,
which is due to the strong constructive interference between
different terms in this region.
Noteworthy also is the large contribution coming from the pion pole
term (Fig.1c), which is much greater than that for the
$\gamma N\to\pi N$ amplitude. This observation is, however, of little
consequence, since in physically realizable case of a bound delta,
where the exchanged pion is far of its mass-shell, the
corresponding contribution turns out to be substantially reduced.
The total cross section shown in Fig.2b has the shape peculiar to
the exothermic reactions with divergence at the zero photon energy.
This is also not the case for a bound delta, which, being
far off-shell, does not produce any real pions, when $E_\gamma~\to$~0.
Thus, we conclude that the single particle cross section will be strongly
affected, when implementing the elementary amplitude into the nucleus.

We proceed to discuss the application of our model for the
$\gamma\Delta^{++}\to\pi^+p$ process
to the nuclear reactions $A(\gamma,\pi^+p)$.
In order to connect all elementary ingredients with $\Delta$'s
dynamics in nuclear medium, we employ here the standart impulse
approximation as well as the closure relation
$\sum_{f}|f \rangle\langle f|\,=\,1$
to sum over the states of residual nucleus.
This approach yields for the lab cross section
\begin{equation}\label{tripl}
\frac{d^3\sigma}{dT_\pi d\Omega_\pi d\Omega_p}=
\frac{M_fM_N\:q\,p_N\:f_\pi(T_\pi)\,f_p(T_p)}
{4(2\pi)^5E_\gamma\left|E_f+E_N(1-\vec{p}_N\cdot(\vec{k}-
\vec{q}\,)/p_N^2)\right|}\:\rho_{\Delta^{++}}(\vec{p}_\Delta\,)\:
\frac{1}{2}\sum_{\lambda=\pm 1}\overline{|t_\lambda|}^2\,.
\end{equation}
Here $T_\pi$ and $T_p$ stand for the pion and proton kinetic energy.
The residual nuclear system is assumed to be $^{11}Be$(g.s.)
with the mass $M_f$ and the total energy $E_f$.
The function $\rho_{\Delta^{++}}(\vec{p}_\Delta\,)$ describes the
distribution of bound deltas in terms of their momentum $\vec{p}_\Delta$.
It is normalised as
\begin{equation}\label{rho}
\int \rho_{\Delta^{++}}(\vec{p}\,)\:\frac{d^3p}{(2\pi)^3}=
\frac{N_\Delta}{4}\,,
\end{equation}
where $N_\Delta$ is the number of deltas in the target nucleus.
In the actual calculation we use
\begin{equation}
\rho_{\Delta^{++}}(\vec{p}_\Delta\,)=4\cdot\frac{4}{3}\pi R^3\:n(p_\Delta)\,,
\end{equation}
where $R$\,=\,3.2\,fm is the square-well radius of $^{12}C$
and the function $n(p_\Delta)$
has a meaning of the $\Delta$'s occupation number inside
the nuclear matter. The factor 4 in (\ref{rho}) accounts for the spin
magnetic number degeneracy. For the $n(p_\Delta)$ we adopt the analysis
of Ref.\cite{Cenni89}, which gives for the $\Delta$ admixture in nuclear
matter the value of the order of 7$\%$ per nucleon.

The  $\Delta$ constituent involved in $\pi^+p$ formation through
the elementary amplitude $t_\lambda$ in (\ref{tripl}) is treated to be
off-shell as determined by the energy and momentum conservation for the
elementary vertex
\begin{eqnarray}
\vec{p}_\Delta &=& \vec{q}+\vec{p}_N-\vec{k}\,,\nonumber\\
E_\Delta &=& E_\pi+E_N-E_\gamma\,.
\end{eqnarray}
It is significant, that the photon energy
240\,MeV\,$\leq E_\gamma\leq$\,400\,MeV considered in this work
is in great part not enough to overcome
the $\Delta$ binding energy ($E_B\approx$\,300\,MeV). Consecuently,
the $A(\gamma,\pi^+p)$ reactions discussed here
are essentially the $\Delta$ decay in nuclear
environment, rather than real $\Delta$ knock-out processes \cite{Gera69}.

In order to take the absorption of emerging particles while
propogating in nuclear medium into account,
the attenuation factors $f_\pi(T_\pi)$ and $f_p(T_p)$
are inserted in (\ref{tripl}).
Assuming square-well
approximation for the $\pi^+$-nucleus and $p$-nucleus optical
potential, these factors have a simple analytical form \cite{Laget72}
\begin{equation}
f_\alpha(T_\alpha)=
\frac{3l_\alpha}{4R}\left[1-\frac{l_\alpha^2}{2R^2}\left\{
1-\left(1+\frac{2R}{l_\alpha}\right)e^{-\frac{2R}{l_\alpha}}\right\}\right]\,,
\quad (\alpha=\pi^+,\,p)\,,
\end{equation}
which is entirely determined by the mean free path $l_\alpha(T_\alpha)$
of the corresponding particle in nuclear matter.

Now we compare our numerical results to the available
data \cite{McKenz97}. Following the experimental conditions of
\cite{McKenz97}, the triple differential cross section
was averaged over the angles $50^o\leq\theta_\pi\leq 130^o$ and
$10^o\leq\theta_p\leq 150^o$.
It is seen from Fig.3  that our results
for the $^{12}C(\gamma,\pi^+p)$ cross section
progressively underestimate the data with increasing photon energy.
In Fig.4, where we show our calculation for the pion kinetic
energy distribution, theoretical predictions agree roughly with the
experimental cross section, when $T_\pi\geq$\,80\,MeV.  At that time,
we can not reproduce relatively large values of the low-energy data.

We would like to note that our model yields a rather essential
fraction of observed pion spectrum, especially for large energies.
The calculated cross section has the same order of magnitude as that
obtained within the VM approach, which attributes all $\pi^+p$-production
events to the final state rescattering processes as mentioned previously.
The results of VM calculation are also shown in Fig.4.
One observes a strong
difference in shape of pion energy spectrum predicted by two models.
Whereas the VM cross section, which is expected to be governed
mainly by the reaction phase space, is concentrated in the region of
small $T_\pi$ and falls off rapidly with increasing pion energy, our curve
peaks at $T_\pi$\,=\,90 MeV and shows only a little value of the
low-energy spectrum.

It is necessary to give some remarks in respect to the validity
of our calculation.
As far as we are concerned with the averaged values
of the exclusive cross section, our results are expected to be not very
sensitive to the choice of the model ingredients, such as the shape of
$\Delta$'s momentum distribution as well as probably
more accurate inclusion of pion and proton distortion in the final state,
etc. Thus, we hope that possible changes brought about by more sophysticated
models will hardly be significant.
However, some effects can markedly modify our quantitative conclusions.

First there is a large uncertainity in the choice of $\Delta$
magnetic moment $\mu_{\Delta^{++}}$.
The value $\mu_{\Delta^{++}}$\,=\,--2.3 n.m.\,\cite{Chao74} used here
differs greatly from $\mu_{\Delta^{++}}$\,=\,5.6 n.m.\ predicted by
simple SU(6) model. One can also mention relatively large value
$\mu_{\Delta^{++}}$\,=\,4.2$\pm$0.5 n.m.\ obtained from the
$\pi^+p\to\pi^+p\gamma$ analysis \cite{Lin91}.
Regarding $\mu_{\Delta^{++}}$ as free parameter,
we obtain a strong increase of the cross section by about a factor of 5,
when varying the magnetic moment in the range from --2.3 to 5.6 n.m.
Thus, it is very important to know the precise value of
$\mu_{\Delta^{++}}$ for an accurate calculation.

Another probable modifications will come from the more
refined study of the role of $\Delta$ isobar configurations in $^{12}C$
ground state. Indeed, the number of deltas of about 0.07 per nucleon
given by the model \cite{Cenni89}, which we adopt here,
is relatively high as compared to $N_\Delta\approx 3.7\%$ \cite{Day76}
for the nuclear matter as well as to those predicted for finite nuclei
(generally less than 4$\%$ \cite{Aren78}).

In conclusion, the aim of this paper is to attract the attention to the
$A(\gamma,\pi^+p)$ reactions, which offer unique possibilities for
testing and extending our knowledge about $\Delta$ degrees of freedom
in nuclei. Our results for $^{12}C(\gamma,\pi^+p)$ reaction are in
rough agreement with the existing data \cite{McKenz97}.
The calculated cross section is comparable to that given
by the FSI mechanism regarded by VM model \cite{Oset92}.
Coming back to the Fig.4, one can speculate that the observed
cross section is filled by both channels, although we recall that the
uncertainity in the model parameters makes our quantitative conclusion
rather ambiguous.
In this respect, it is important to point out that more usefull information
on the role of virtual deltas in $A(\gamma,\pi^+p)$ reactions
may be obtained from the true exclusive data (not averaged
over a wide range of angles as presented in \cite{McKenz97}).
Such measurements make it possible to separate the FSI background
from the $\Delta^{++}(\gamma,\pi^+p)$ events. Experiments along these
lines are in progress at Tomsk synchrotron.
To complete this discussion, we would like to note that
the $A(\gamma,\pi^+p)$ measurements on lightest nuclei,
such as deuteron and $^3H$, are also highly desirable.
In this case, the FSI contribution is assumed to be
of minor importance and the information about $\Delta^{++}(\gamma,\pi^+p)$
amplitude may be interpreted unambiguously.
What is more, minimal number of nucleons allows a carefull
microscopic treatment of these processes.

This work was supported in part by the Russian Foundation for Fundamental
Research under Contract No.\ 96-02 16742-a.

\newpage
\begin{center}
\bf Figure captions
\end{center}

\begin{itemize}
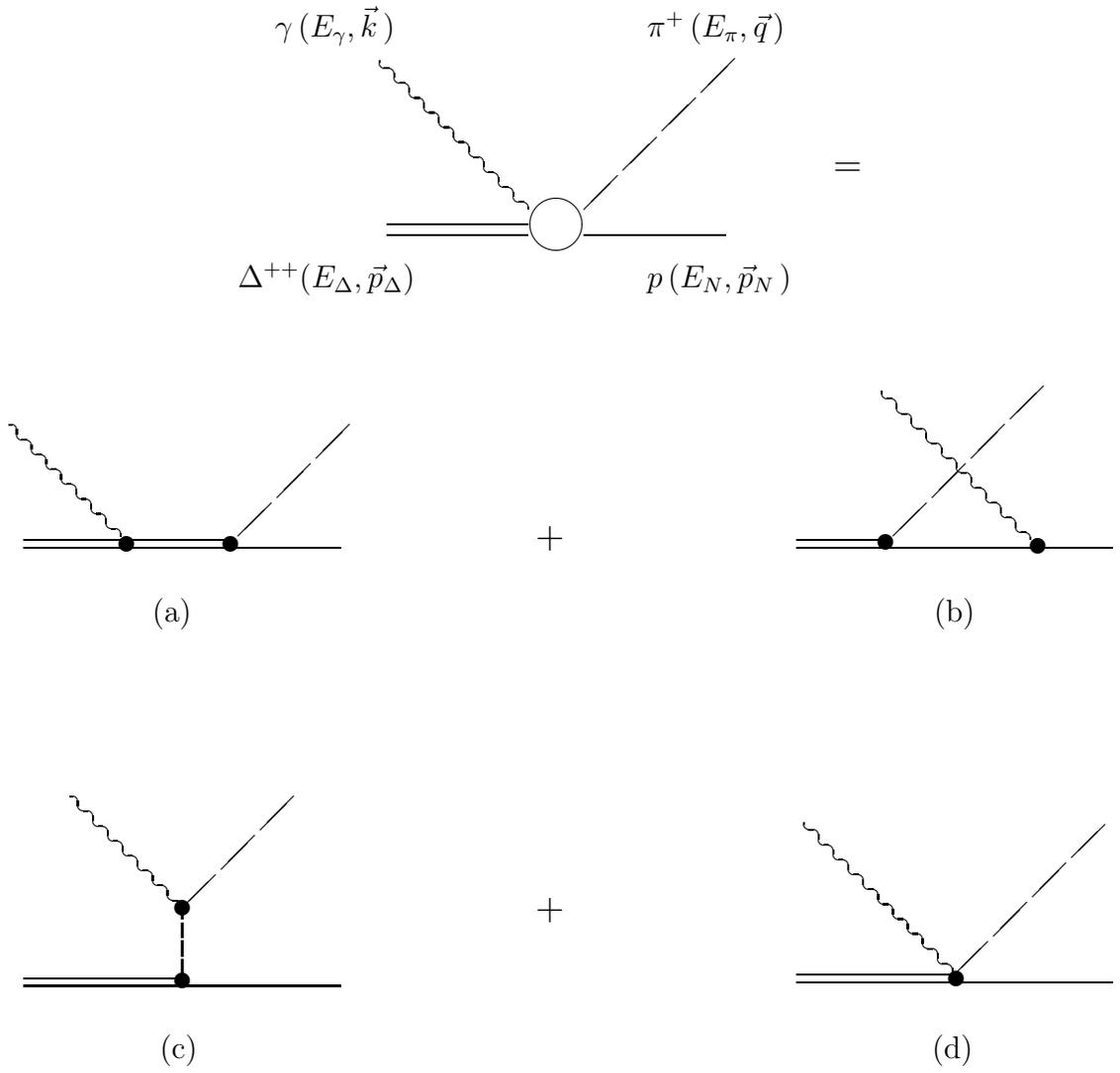
\item[\bf FIG.1]
Diagrams for the $\gamma\Delta^{++}\to\pi^+p$
amplitude used in the present calculation:
s-channel term (a),\
u-channel term (b),\
pion pole term (c),\
seagull term (d).

\item[\bf FIG.2]
(a) Angular distribution for the $\gamma\Delta^{++}\to\pi^+p$ reaction
in c.m.\ system. The curves present the contributions from different
terms depicted if Fig.1:
full calculation (solid),\
s-channel term (dashed),\
u-channel term (dash-dotted),\
pion pole term (dotted),\
seagull term (dash-double dotted).\\
(b) Total cross section for the $\gamma\Delta^{++}\to\pi^+p$ reaction.

\item[\bf FIG.3]
Triple differential cross section for the $^{12}C(\gamma,\pi^+p)$
reaction versus photon energy, averaged over the angles of emitted
particles $50^o\leq\theta_\pi\leq 130^o$ and
$10^o\leq\theta_p\leq 150^o$. The pion and proton thresholds are set
at $T_\pi$\,=\,30\,MeV and $T_p$\,=\,50\,MeV.
The data are taken from \cite{McKenz97}.

\item[\bf FIG.4]
Pion kinetic energy distribution for the
$^{12}C(\gamma,\pi^+p)$ reaction.
Step line is the VM calculation
\cite{Oset92}. The data are from \cite{McKenz97}.

\end{itemize}

\newpage
\unitlength 0.25mm
\begin{center}
\begin{picture}(800,700)(0,0)

\multiput(240,602)(8,-8){10}{\oval(8,8)[lb]}
\multiput(240,594)(8,-8){10}{\oval(8,8)[rt]}
\put(240,514){\line(1,0){76}}
\put(240,508){\line(1,0){76}}
\multiput(346,522)(21,21){4}{\line(1,1){18}}
\put(346,508){\line(1,0){76}}
\put(331,514){\circle{30}}
\put(180,615){$\gamma\,(E_\gamma,\vec{k}\,)$}
\put(380,615){$\pi^+\,(E_\pi,\vec{q}\,)$}
\put(160,480){$\Delta^{++}(E_\Delta,\vec{p}_\Delta)$}
\put(380,480){$p\,(E_N,\vec{p}_N\,)$}
\put(480,540){\Large =}

\multiput(45,402)(8,-8){7}{\oval(8,8)[lb]}
\multiput(37,402)(8,-8){8}{\oval(8,8)[rt]}
\put(45,344){\line(1,0){112}}
\put(45,340){\line(1,0){170}}
\multiput(160,346)(21,21){3}{\line(1,1){18}}
\put(100,342){\circle*{8}}
\put(156,342){\circle*{8}}
\put(114,300){(a)}
\put(320,340){\Large +}

\multiput(510,424)(8,-8){10}{\oval(8,8)[lb]}
\multiput(510,416)(8,-8){10}{\oval(8,8)[rt]}
\multiput(512,346)(21,21){4}{\line(1,1){18}}
\put(460,344){\line(1,0){50}}
\put(460,340){\line(1,0){170}}
\put(508,343){\circle*{8}}
\put(590,341){\circle*{8}}
\put(534,300){(b)}

\multiput(78,202)(8,-8){7}{\oval(8,8)[lb]}
\multiput(70,202)(8,-8){8}{\oval(8,8)[rt]}
\multiput(130,146)(21,21){3}{\line(1,1){18}}
\put(45,108){\line(1,0){85}}
\put(45,104){\line(1,0){170}}
\multiput(130,110)(0,10){4}{\line(0,1){8}}
\put(130,107){\circle*{8}}
\put(130,146){\circle*{8}}
\put(320,140){\Large +}
\put(118,65){(c)}

\multiput(468,192)(8,-8){10}{\oval(8,8)[lb]}
\multiput(468,184)(8,-8){10}{\oval(8,8)[rt]}
\multiput(545,110)(21,21){4}{\line(1,1){18}}
\put(460,110){\line(1,0){85}}
\put(460,106){\line(1,0){170}}
\put(546,108){\circle*{8}}
\put(533,65){(d)}

\end{picture} \\

\Large Figure 1.
\end{center}


\newpage
\begin{thebibliography}{99}

\bibitem{Green76}
A.M.\,Green, Rep.Progr.Phys.\ {\bf 39} (1976) 1109

\bibitem{Aren78}
H.J.\,Weber, H.\,Arenh\"ovel, Phys.Rep.\ {\bf 36} (1978) 277

\bibitem{Gera69}
S.B.\,Gerasimov, ZhETF Pis.Red. {\bf 14} (1971) 385
(JETP Lett.\ 14 (1971) 260)

\bibitem{McKenz97}
M.\,Liang, D.\,Branford, T.\,Davinson {\it et al}., Phys.Lett.\
{\bf B411} (1997) 244

\bibitem{Oset92}
R.\,Carrasco, E.\,Oset, Nucl.Phys.\ {A536} (1992) 445

\bibitem{Chao74}
Y.A.\,Chao, R.K.P.\,Zia, Nuovo Cimento {\bf A19} (1974) 651

\bibitem{CGLN}
G.F.\,Chew, M.L.\,Goldberger, F.E.\,Low, Y.\,Nambu,  Phys.\,Rev.\ {\bf 106}
(1957) 1345

\bibitem{Cenni89}
R.\,Cenni, F.\,Conte, U.\,Lorenzini, Phys.Rev.\ {\bf C39} (1989) 1588

\bibitem{Laget72}
J.-M.\,Laget, Nucl.Phys.\ {\bf A194} (1972) 81

\bibitem{Day76}
B.D.\,Day, R.\,Coester, Phys.\,Rev.\ {\bf C13} (1976) 1720

\bibitem{Lin91}
D.\,Lin, R.K.\,Liou, Phys.\,Rev.\ {\bf C43} (1991) 930
\end{thebibliography}
\end{document}